\documentclass{aa} 
\usepackage[utf8]{inputenc}
\usepackage[english]{babel}
\usepackage{mathtools}
\usepackage{amsfonts}
\usepackage{amssymb}
\usepackage{graphicx}
\usepackage{hyperref}
\hypersetup{colorlinks = true,
            allcolors = {blue}}

\usepackage{natbib}
\setcitestyle{round}

\usepackage{caption}
\usepackage{subcaption}
\usepackage[T1]{fontenc}  

\newcommand{\beq}{\beq}
\newcommand{\eeq}{\eeq}
\newcommand{\beqa}{\begin{eqnarray}}
\newcommand{\eeqa}{\end{eqnarray}}

\begin{document}

\title{Spatial variation of periods of ion and neutral waves in a solar magnetic arcade}
\titlerunning{Two-fluid waves in a magnetic arcade}
\authorrunning{B. Ku\'zma et al.}
\offprints{B. Ku\'zma}

\author{{B.~Ku\'zma \inst{\ref{inst1},\ref{inst2}}}  \and {K.~Murawski\inst{\ref{inst2}}} \and {Z.E.~Musielak\inst{\ref{inst3},\ref{inst4}}} \and 
{S.~Poedts \inst{\ref{inst1},\ref{inst2}}}  \and
{D.~W\'ojcik \inst{\ref{inst2}}}}

\institute{
    Center for Mathematical Plasma Astrophysics, Department of Mathematics, KU Leuven, Celestijnenlaan 200B, 3001 Leuven, Belgium \label{inst1}
    \and Institute of Physics, University of Maria Curie-Sk{\l}odowska, Pl. Marii Curie-Sk\l{}odowskiej 5, 20-031 Lublin, Poland \label{inst2}
	\and Department of Physics, University of Texas at Arlington, Arlington, TX 76019, USA \label{inst3}
	\and Leibniz-Institut f\"ur Sonnenphysik (KIS), Sch\"oneckstr. 6, 79104 Freiburg, Germany	 \label{inst4}
   }
\date{Received; accepted}

%

%


\abstract
{We present a new insight into the propagation of ion magnetoacoustic and neutral acoustic waves in a magnetic arcade in the lower solar atmosphere.}
{By means of numerical simulations, we aim to: (a)~study two-fluid waves propagating in a magnetic arcade embedded in the partially-ionized, lower solar atmosphere; and (b)~investigate the impact of the background magnetic field configuration on the observed wave-periods.} 
{We consider a 2D approximation of the gravitationally stratified and partially-ionized lower solar atmosphere consisting of ion + electron and neutral fluids that are coupled by ion-neutral collisions.  In this model, the convection below the photosphere is responsible for the excitation of ion magnetoacoustic-gravity and neutral acoustic-gravity waves.  }
{We find that in the solar photosphere, where ions and neutrals are strongly coupled by collisions, ion magnetoacoustic-gravity and neutral acoustic-gravity waves have periods ranging from $250\;$s to $350\;$s. 
In the chromosphere, where the collisional coupling is weak, the wave characteristics strongly depend on the magnetic field configuration. 
Above the foot-points of the considered arcade, the plasma is dominated by a vertical magnetic field along which ion magnetoacoustic-gravity waves propagate. These waves exhibit a broad range of periods with the most prominent periods of $180\;$s, $220\;$s, and $300\;$s. 
Above the main loop of the solar arcade, where mostly horizontal magnetic field lines guide ion magnetoacoustic-gravity waves, the main spectral power reduces to the period of about $180\;$s and longer wave-periods do not exist. 
} 
{In photospheric regions, ongoing solar granulation excites a broad spectrum of wave-periods that undergoes complex interactions: mode-coupling, refractions through the inhomogeneous atmosphere, real physical absorption and conversion of wave power. We found that, in addition, the magnetic arcade configuration with a partially-ionized plasma drastically changes the image of wave-periods observed in the upper layers of the chromosphere and corona. 
Our results are in agreement with the recent observational data reported by Wi\'sniewska et al.\ (2016) and Kayshap et al.\ (2018).}



\maketitle

\section{Introduction}
Plasma consisting of two fluids, namely ions + electrons and neutrals, that is in a gravity-free and homogeneous equilibrium state, permeated by a uniform magnetic field, can guide slow and fast ion magnetoacoustic waves (MAWs), and neutral acoustic waves, Alfv\'en waves and entropy/thermal modes \citep[e.g.][]{Nakariakov2005, Ballester2018}. 
Small-amplitude Alfv\'en waves directly alter azimuthal (tangential-perpendicular) components of the ion velocity and magnetic field, while the other plasma quantities remain essentially unperturbed. 
Ion and neutral entropy/thermal modes correspond to non-propagating mass density perturbations \citep[e.g.][]{Goedbloed_2004,Murawski2011}. 
Neutral acoustic waves are coupled to ion MAWs through ion-neutral collisions, and these waves are dispersive and collisions introduce characteristic spatial and temporal scales \citep[e.g.][]{Zaqarashvili2011, Ballester2018}. It is the subject of some on-going discussions whether two-fluid effects are negligible at low frequency or not (e.g.\ \citealp{Soler2013}, \citealp{Ballai2019}, among others). Recently published results showed, however, that two-fluid effects play an important role even in the case of low frequency waves when non-linear effects occur, especially in the context of wave damping and plasma heating (e.g.\ \citealp{Kuzma2019}, \citealp{Wojcik2020}, \citealp{Murawski2020}, \citealp{Fan2020}). 

In a gravitationally-stratified atmosphere, internal gravity (henceforth 'gravity') waves join the above-mentioned waves. These gravity waves are unable to propagate along the vertical direction but otherwise they couple to acoustic waves or MAWs and create a hybrid entity named magnetoacoustic-gravity (MAG) waves \citep[e.g.][]{Vigeesh2017}. 
In the extreme limit of a magnetic-free atmosphere, the magnetic (slow MAWs and Alfv\'en) waves disappear and fast MAWs become purely acoustic \citep[e.g.][]{Nakariakov2005}.

The solar granulation, which operates within the lowest atmospheric layer, called the photosphere, is a source of various dynamical events such as eddies, downdrafts, and the above-described waves \citep[e.g.][]{Vigeesh2017}. 
In a quiet photospheric region, the waves spectrum exhibits its main power around the period of $300\;$s \citep[e.g.][]{Leighton1962, ChristensenDalsgaard1991}, and ions and neutrals are strongly coupled there by frequent collisions. 
However, in a magnetic flux-tube and in the chromosphere the average period of the oscillations is close to $180\;$s \cite[e.g.][]{DeubnerFleck1990}. 
As a result, the ion-neutral collision frequency falls off with height and the ion-neutral coupling becomes weak in the upper chromosphere. 

There were a number of attempts to explain the origin of the so-called "$3$-min oscillations" \cite[e.g.][]{Lamb1909,Lamb1911,Moore1964,Moore1972,Ulmschneider1978,Cuntz1998,Fawzy2002,Musielak2006,Fawzy2012,Routh2014,Kraskiewicz_2019}. 
In particular, \cite{Fleck1993} compared four different spectra of injected waves at the lower boundary. They found that a shock overtaking mechanism removes high frequency waves, while essentially any driver results in $3$-min chromospheric oscillations as a property of the stratification.
Their results were confirmed within the framework of magnetohydrodynamic (MHD) and a two-fluid model by  \cite{Kraskiewicz_2019} and \cite{Wojcik2018a},  respectively, who additionally showed that ions and neutrals oscillate differently, while being excited by a monochromatic driver in a magnetic-free atmosphere. 
A strong magnetic field case was discussed by \cite{2011ApJ...728...84B} who observed three-minute oscillations in the chromosphere above sunspot umbrae and developed a model based on the ideal MHD equations with a uniform vertical background magnetic field and a temperature profile corresponding to sunspot atmospheres. 
The large wave-periods in the spectrum of the broadband pulse were filtered out so that wave-periods below the acoustic cutoff wave-period resonated inside the chromospheric cavity with about $3$-min. 

Recently, \cite{Wojcik2018a} performed two-fluid numerical simulations of a partially-ionized solar atmosphere that is permeated by a weak and initially vertical magnetic field, and concluded that MAWs that are excited by granulation, convert their main power from the period of about $300\;$s in the photosphere into oscillations of about $220\;$s in the chromosphere. 

The main goal of the present paper is to generalise the two-fluid models presented by \cite{Wojcik2018a} and \cite{Wojcik2019} to a magnetic arcade that contains both regions of a weak horizontal magnetic field which interacts with ion MAG waves, and two zones of stronger, mostly vertical magnetic field that acts like a wave guide. The obtained results discover the role of waves in transporting energy from the solar photosphere to the upper layers of the solar chromosphere, heating the latter and the effects of cutoff frequencies on the propagation of these waves. 
The results of our numerical simulations are compared to observational data given by \cite{2016ApJ...819L..23W} and \cite{Kayshap2018}, and good agreement was found.

In the following section  the two-fluid equations are presented and discussed. 
Section 3 contains a presentation of the results of the numerical simulations and in Section 4 the main results are summarised. 

\section{Two-fluid equations}
We here use the two-fluid equations in which (ion $_{\rm i}$ and neutral $_{\rm n}$) mass densities, 
$\varrho_{\rm i,n}$, 
velocities, ${\bf V}_{\rm i,n}$, (ion+electron $_{\rm ie}$ and neutral $_{\rm n}$) gas pressures, 
$p_{\rm i \, e,n}$, and magnetic field ${\bf B}$ evolve according to (e.g. \citealp{Oliver2016,Ballester2018})
\begin{equation}
\frac{\partial \varrho_{\rm n}}{\partial t}+\nabla\cdot(\varrho_{\rm n} \mathbf{V}_{\rm n}) = 0 \,,
\label{eq:neutral_continuity}
\end{equation}
\begin{equation}
\frac{\partial \left( \varrho_{\rm n} \mathbf{V}_{\rm n} \right)}{\partial t}+\nabla \cdot (\varrho_{\rm n} \mathbf{V}_{\rm n} \mathbf{V}_{\rm n}+p_{\rm n} \mathbf{I}) = \alpha_{\rm in}({\bf V}_{\rm i}-{\bf V}_{\rm n}) + \varrho_{\rm n} \mathbf{g},
\label{eq:neutral_momentum}
\end{equation}
\begin{equation}
\begin{split}
\frac{\partial E_{\rm n}}{\partial t}+\nabla\cdot[(E_{\rm n}+p_{\rm n})\mathbf{V}_{\rm n}] = 
\alpha_{\rm in}\mathbf{V}_{\rm n}\cdot({\bf V}_{\rm i}-{\bf V}_{\rm n})\\ +Q_{n} ^{in} 
+ \varrho_{\rm n} \mathbf{g} \cdot \mathbf{V}_{\rm n},
\end{split}
\label{eq:neutral_energy}
\end{equation}
where the heat production and exchange term for neutrals is given by
\begin{equation}
Q_{n}^{in} = \alpha_{\rm in} \left[\frac{1}{2} |{\mathbf V}_{\rm i}-{\mathbf V}_{\rm n} |^2 + \frac{3k_{\rm B}}{m_{\rm H}(\mu_{\rm i}+\mu_{\rm n})}\left(T_{\rm i}-T_{\rm n} \right )\right],
\end{equation}
and the neutral (internal + kinetic) energy density is defined as
\begin{equation}
E_{\rm n} = \frac{p_{\rm n}}{\gamma-1} + \frac{\varrho_{\rm n}|\mathbf{V}_{\rm n}|^2}{2}.
\end{equation}
For ions+electrons mixture we use the following set of MHD equations: 
\begin{equation}
\frac{\partial \varrho_{\rm i}}{\partial t}+\nabla\cdot(\varrho_{\rm i} \mathbf{V}_{\rm i}) = 0,
\label{eq:ion_continuity}
\end{equation}
\begin{equation}
\begin{split}
\frac{\partial \left( \varrho_{\rm i} \mathbf{V}_{\rm i} \right)}{\partial t}+\nabla \cdot (\varrho_{\rm i} \mathbf{V}_{\rm i} \mathbf{V}_{\rm i}+p_{\rm ie} \mathbf{I}) = \frac{1}{\mu_0} \left(\nabla \times \mathbf{B}\right) \times \mathbf{B} + \\ \alpha_{\rm in}({\bf V}_{\rm n}-{\bf V}_{\rm i}) +\varrho_{\rm i} \mathbf{g},
\end{split}
\label{eq:ion_momentum}
\end{equation}
\begin{equation}
\frac{\partial \mathbf{B}}{\partial t} = \nabla \times (\mathbf{V_{\rm i} \times }\mathbf{B}), \hspace{0.5cm} \nabla \cdot \mathbf{B}=0,
\label{eq:ions_induction}
\end{equation}
\begin{equation}
\begin{split}
\frac{\partial E_{\rm i}}{\partial t}+\nabla\cdot\left[\left(E_{\rm i}+p_{\rm ie} + \frac{|\mathbf{B}|^2}{2\mu_0}\right)\mathbf{V}_{\rm i}-
\frac{1}{\mu_0}\mathbf{B}(\mathbf{V_{\rm i}}\cdot \mathbf{B})\right] = \\ 
\alpha_{\rm in}\mathbf{V}_{\rm i}\cdot({\bf V}_{\rm n}-{\bf V}_{\rm i}) + Q_{i}^{in} 
+ \varrho_{\rm i} \mathbf{g}  \cdot \mathbf{V}_{\rm i}+L_{\rm r}, 
\end{split}
\label{eq:ion_energy}
\end{equation}
with a similar heat exchange and production term for ions given by
\begin{equation}
Q_{i}^{in} = \alpha_{\rm in} \left[\frac{1}{2} |{\mathbf V}_{\rm i}-{\mathbf V}_{\rm n} |^2 + \frac{3k_{\rm B}}{m_{\rm H}(\mu_{\rm i}+\mu_{\rm n})}\left(T_{\rm n}-T_{\rm i} \right )\right],
\end{equation}
and the ion (internal + kinetic + magnetic) energy density defined as
\begin{equation}
E_{\rm i} = \frac{p_{\rm ie}}{\gamma-1} + \frac{\varrho_{\rm i}|\mathbf{V}_{\rm i}|^2}{2} + \frac{|\mathbf{B}|^2}{2\mu_0} .
\end{equation}
Here, ${\bf g} = [0, -g, 0]$ is a gravity vector with its magnitude $g = 274.78$ m s$^{-2}$, $\alpha_{\rm in}$ denotes the coefficient of collisions between ion and neutral particles \citep[e.g.][and references cited therein]{Ballester2018}, $\mu_{\rm i}=0.58$ and $\mu_{\rm n}=1.21$ are the mean masses of ions and neutrals, respectively, which are specified by the OPAL solar abundance model, 
while $m_{\rm H}$ is the hydrogen mass, $k_{\rm B}$ is the Boltzmann constant,  
and $\gamma=1.4$ is the specific heats ratio.
Note that \cite{Oliver2016} assumed hydrogen plasma, thus they implied $\mu_{\rm i}\approx\mu_{\rm n}$ in Eqs.~(4) and (10). The symbol $L_{\rm r}$ denotes the radiative loss term, which consists of two separate parts: thick cooling that operates in the low atmospheric layers, and thin cooling that works in the upper atmospheric regions. At every time-step, we calculate the optical depth starting from infinity and ending at the bottom boundary. For optical depths higher than 0.1, we use the thick cooling approximation of \cite{Abbett_2011}. For optical depths lower than 0.1 we use thin cooling \citep{Moore1972}. 
The other symbols in Eqs.~(1)-(11) have their standard meaning. 
The pressures of ions + electrons and neutrals are given by the ideal gas laws as
\begin{equation}
p_n=\frac{k_B}{m_H\mu_{\rm n}}\varrho_{\rm n} T_{\rm n}, \hspace{0.5cm}\hbox{and}\hspace{0.5cm} p_{\rm ie}=\frac{k_B}{m_H\mu_{\rm i}}\varrho_{\rm i} T_{\rm i}.
\label{eq:pressures}
\end{equation} 
Having specified the temperature profile and using Eqs.~(1)-(12), we obtain the hydrostatic ion and neutral mass densities, and the 
gas pressure profiles. 
As a result of a small mass of electrons in comparison to ions and neutrals, we neglected all electron-associated terms in the induction equation \citep{Ballester2018} and also all 
terms corresponding to viscosity, magnetic resistivity, and ionization and recombination. These terms and effects will be included in future studies.

\section{Numerical simulations and results}
\subsection{Numerical set up, boundary and initial conditions}
To solve the two-fluid equations numerically, we use the JOANNA code \citep{Wojcik2018a,Wojcik2019} which solves the two-fluid equations in the form (\ref{eq:neutral_continuity})-(\ref{eq:pressures}). 
In our numerical experiments, we set the Courant-Friedrichs-Lewy \citep{Courant1928} number equal to 0.9 and choose a second-order accuracy in space and a four-stage, third-order strong stability preserving Runge-Kutta method \citep{Durran2010} for integration in time, supplemented by adopting the Harten-Lax-van Leer Discontinuities (HLLD) approximate Riemann solver \citep{Miyoshi2010} and the divergence of magnetic field cleaning method of \cite{Dedner2002}.

The two-dimensional simulation box is specified as $-10.24\;$Mm $<x<$ $10.24\;$Mm along the horizontal ($x$-) direction. 
Along the vertical ($y$-) direction the region from $y=-2.56\;$Mm up to $y=7.68\;$Mm is covered by a uniform grid of cell size $\Delta x=\Delta y=20\;$km.
Above this uniform-grid zone, a region of the stretched grid is implied up to $y=60\;$Mm and it is covered by 128 cells. 
With the use of this non-uniform grid any incoming signal is damped close to the upper boundary. It has been found that such stretched grid 
significantly reduces spurious reflections of the incoming signal (e.g.\ \citealp{Kuzma2018}). The size of the 128 stretched numerical cells grows with height. Thus, cells close to the area with uniform grid have a similar size and the transition from a uniform to a fully stretched grid is smooth and does not have any significant impact on propagating waves. 
The implemented grid resolves  all spatial structures well, particularly those associated with solar granulation. 

At the top and bottom boundaries, we fix all plasma quantities, except ion and neutral velocity, to their equilibrium values. Holding them fixed allows us to reduce numerical noise due to the integration of the gravity term as well as numerically induced reflections of the incoming signal. 
At the lateral sides, 
periodic boundary conditions are implemented. The physical system is taken to be invariant along the $z-$direction (with $\partial / \partial z =0$) and $V_z=B_z=0$ was set throughout the whole time. As a result of that, Alfv\'en waves were removed from the system which still allows to propagate MAG waves. Without implementing any transversal velocity perturbation this condition is maintained self-consistently. 



%
\begin{figure}
\begin{center}
\includegraphics[width=1.1\linewidth]{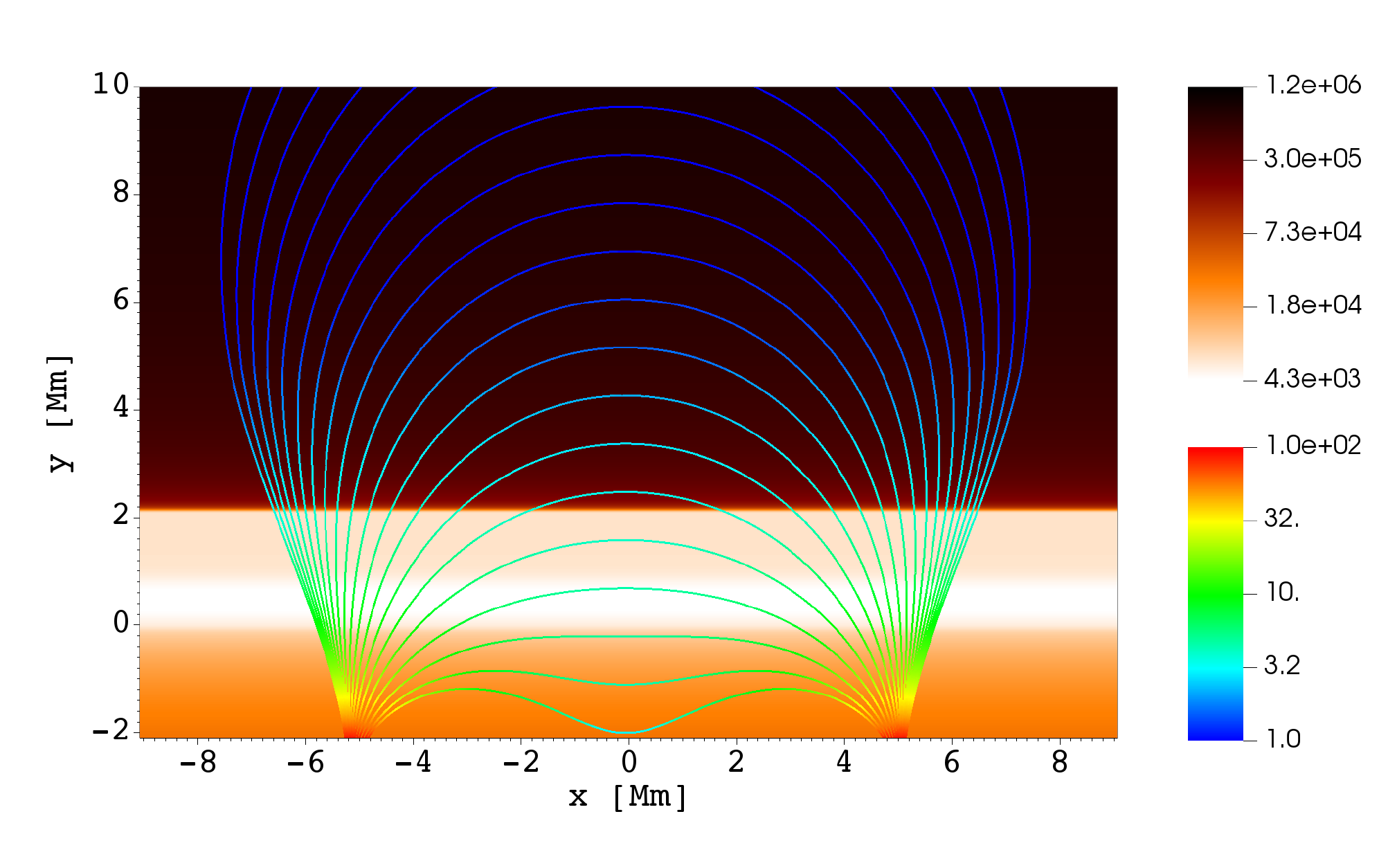}
\includegraphics[width=1.1\linewidth]{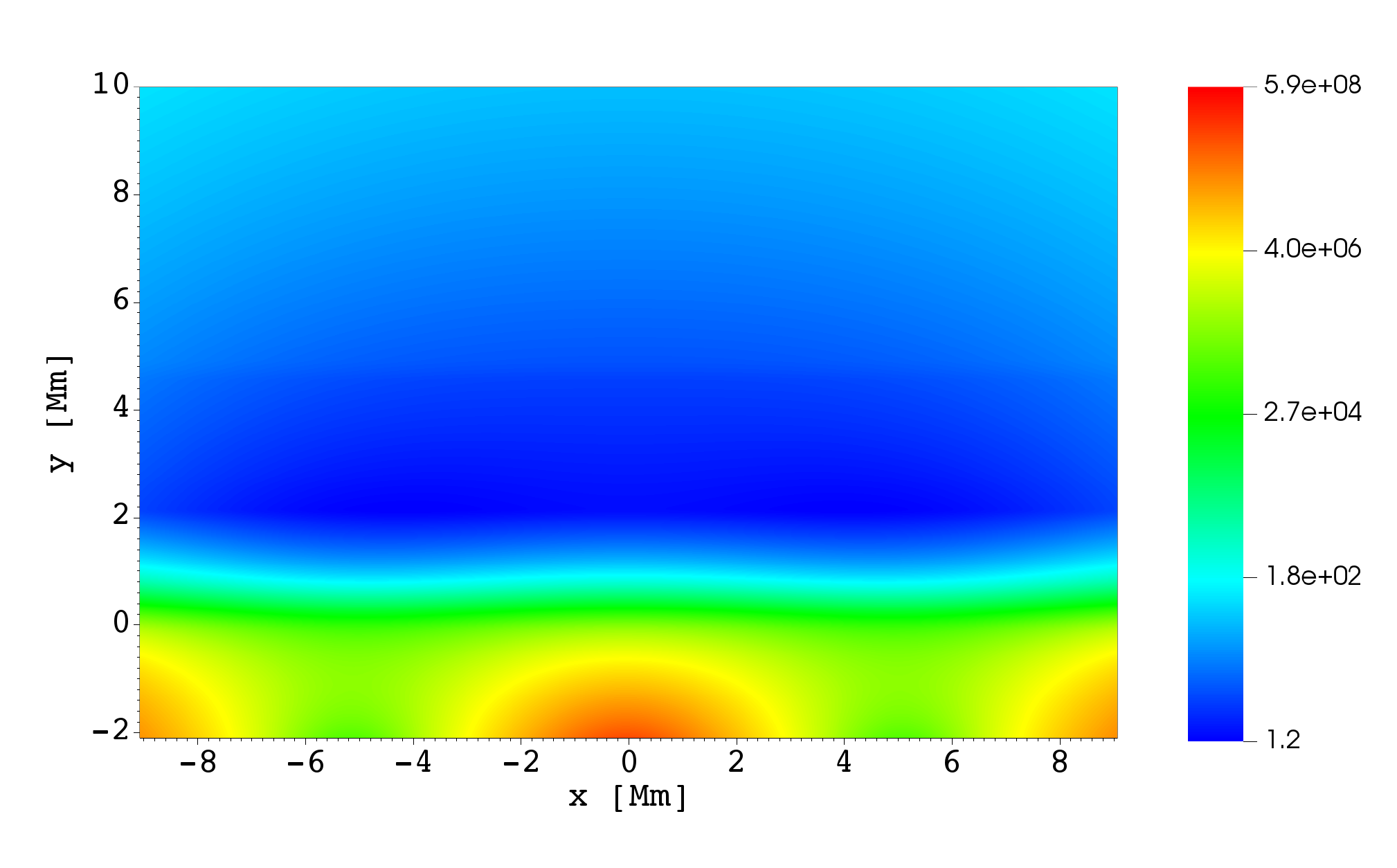}
\end{center}
\vspace{-1cm}
\caption{Initial spatial profiles of logarithm of the ion temperature, $T_{\rm i}$, expressed in K, overlayed by the magnetic field lines with magnitude of the magnetic field, $B$, expressed in G (top) and logarithm of the plasma beta (bottom). 
}

\label{fig:spat_prof}
\end{figure}

\begin{figure*}[ht!]
	\begin{center}
		\mbox{
			\includegraphics[width=0.5\linewidth]{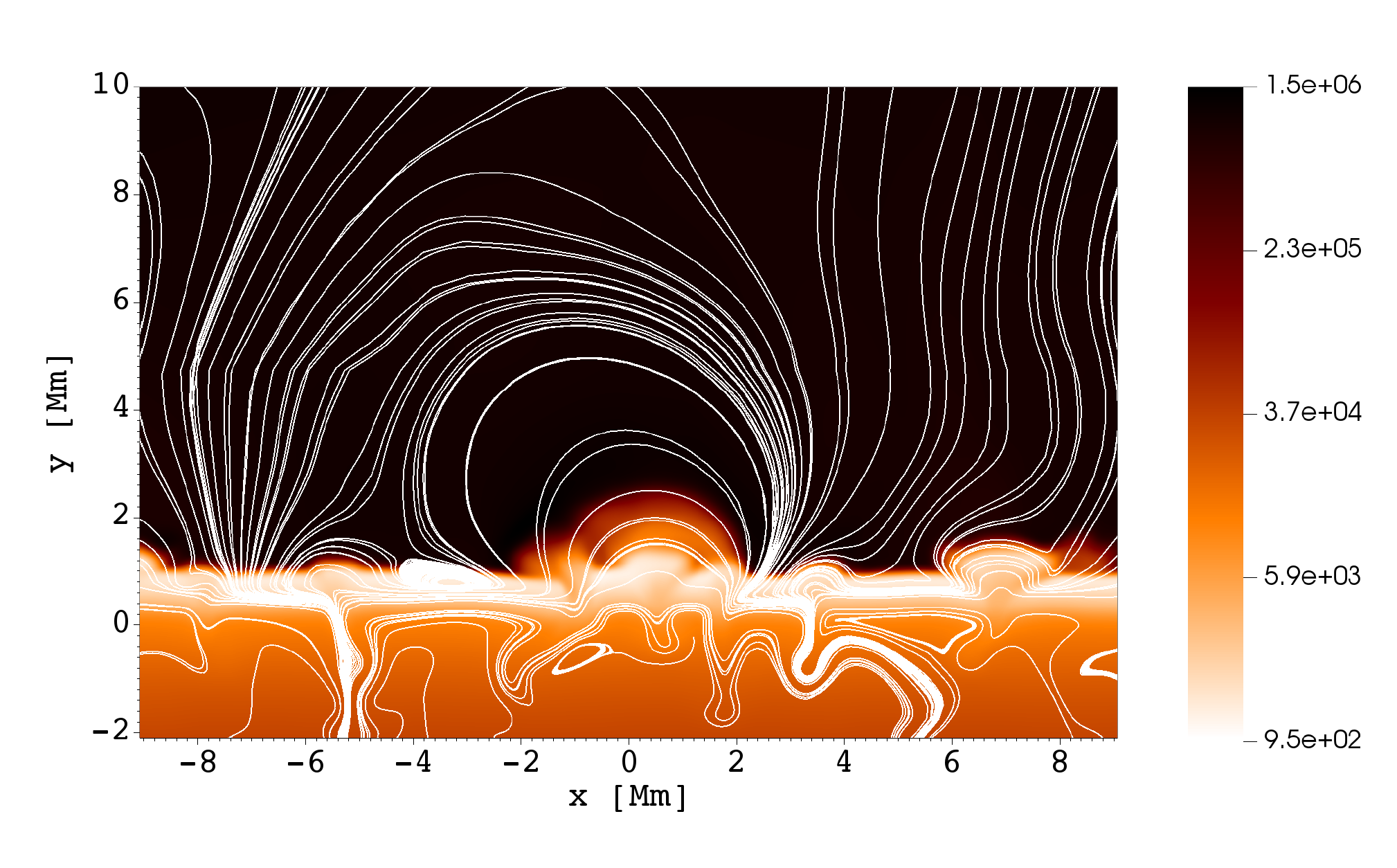}
			\includegraphics[width=0.5\linewidth]{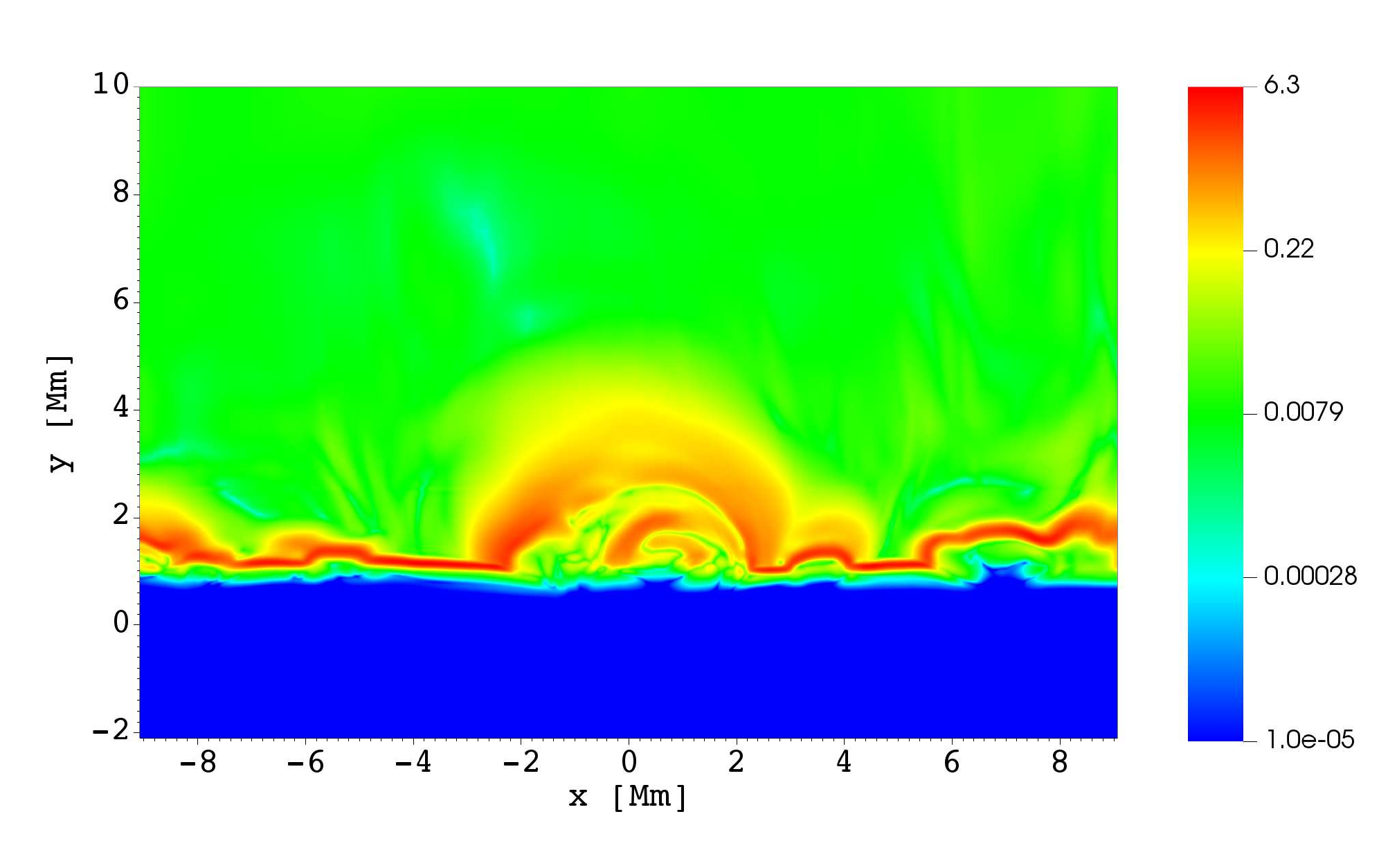}
					}
		\mbox{
			\includegraphics[width=0.5\linewidth]{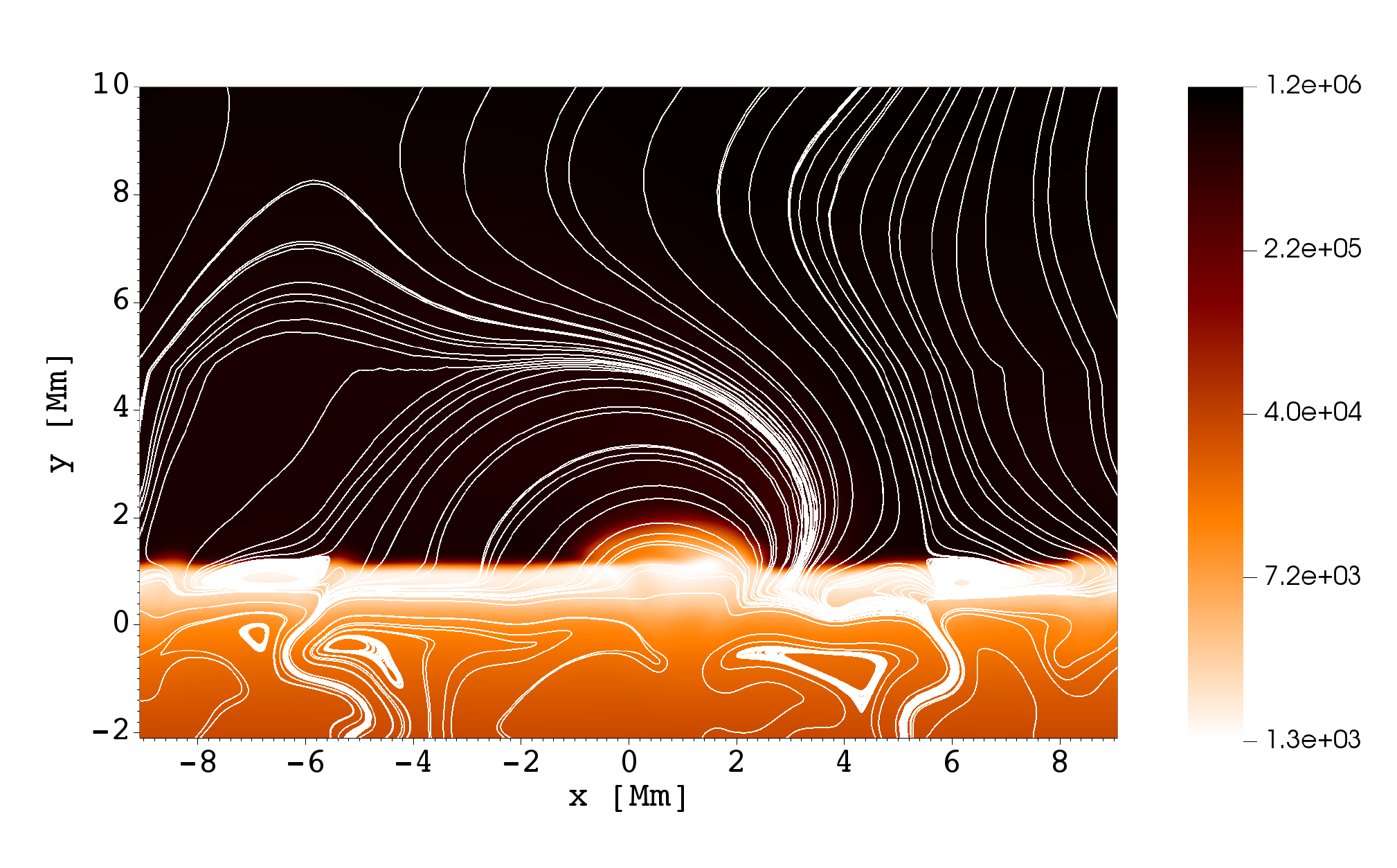}
			\includegraphics[width=0.5\linewidth]{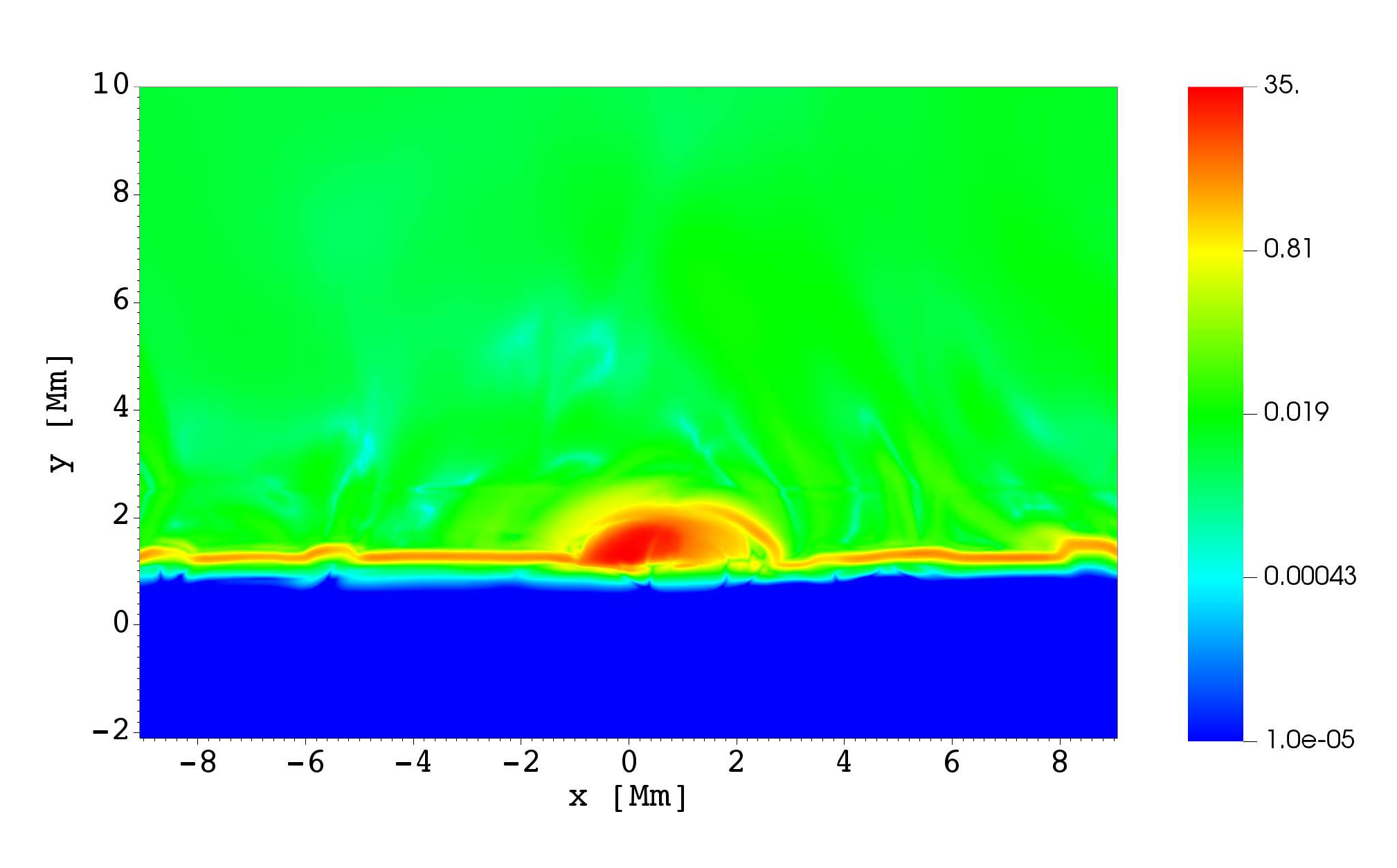}
					}
							\mbox{
			\includegraphics[width=0.5\linewidth]{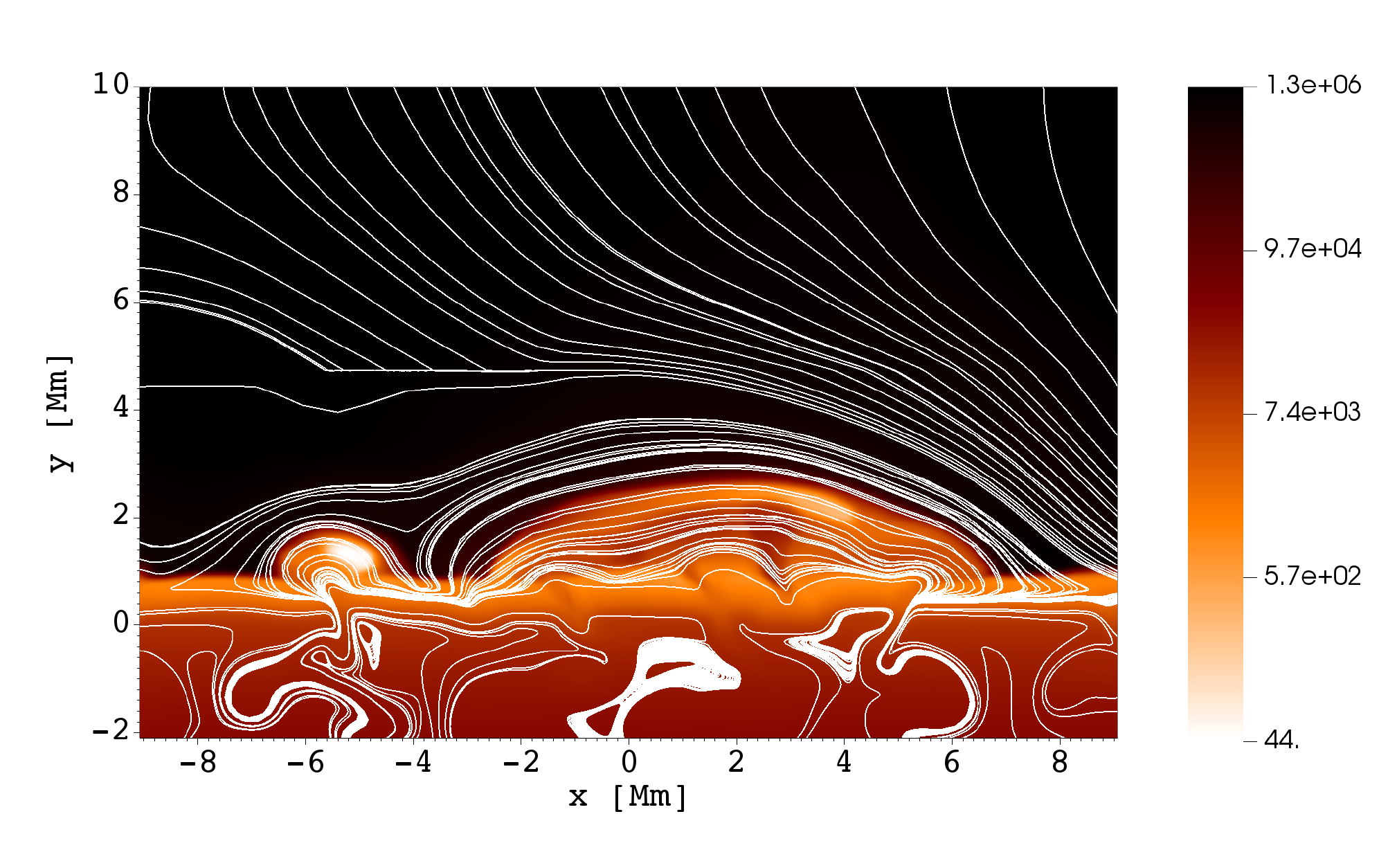}
			\includegraphics[width=0.5\linewidth]{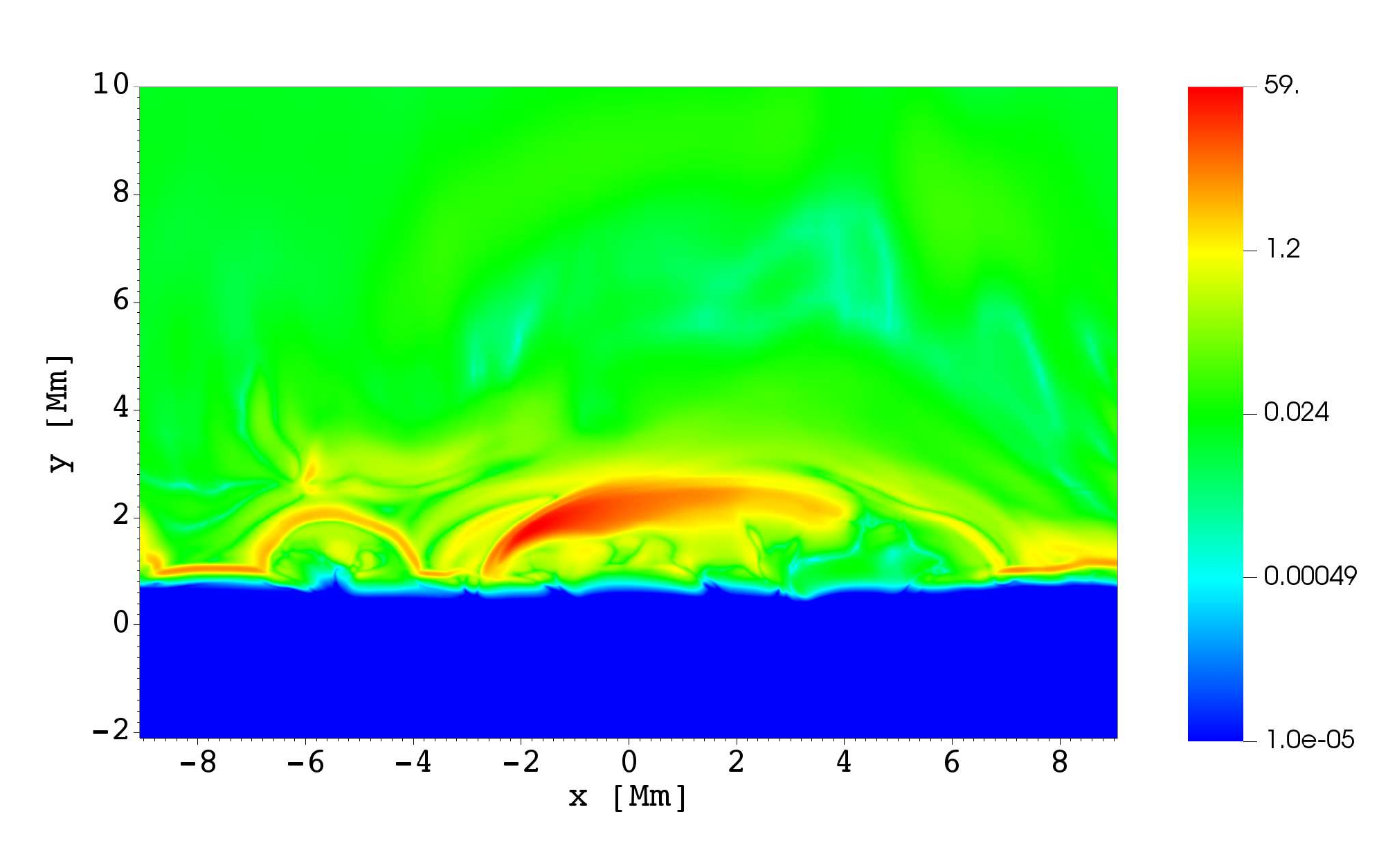}
					}
		\caption{Spatial profiles of logarithm of the ion temperature, $T_{\rm i}$, expressed in K, overlayed by the magnetic field lines (left) and the corresponding ion-neutral drift, $|V_{\rm i}-V_{\rm n}|$, \textbf{expressed in km s}$^{-1}$ (right) at $t=3800$ s, $t=4000$ s, and $t=4200$ s (from top to bottom). 
		}
	\end{center}
	\label{fig:equil_BvecCa}
\end{figure*}

At $t=0\;$s, we set all plasma quantities to their hydrostatic values specified by the temperature given by the semi-empirical model of \cite{Avrett2008} (Fig.~1, top). The equilibrium magnetic field is taken in the form of an arcade with its foot points settled at $x=-5.12\;$Mm and $x=5.12\;$Mm  \citep{Low1985}. The horizontal, $B_{x}$, vertical, $B_y$ and transversal, $B_z$ components of magnetic field are described by:
\begin{eqnarray}
    B_x(x,y)&=\frac{-2S(x-a_1)(y-b)}{((x-a_1)^2+(y-b)^2)^2 } +\frac{-2S(x-a_2)(y-b)}{((x-a_2)^2+(y-b)^2)^2 }\, , \nonumber
    \\
    B_y(x,y)&=\frac{S((x-a_1)^2-(y-b)^2)}{((x-a_1)^2+(y-b)^2)^2}+ \frac{S((x-a_2)^2-(y-b)^2)}{((x-a_2)^2+(y-b)^2)^2}\, ,
    \\
    B_z(x,y)&=0 \, . \nonumber
\end{eqnarray}
Here $a_1=-5.12$ Mm, $a_2=5.12$ Mm and $b=-2.8$ Mm indicate locations of the magnetic singularities, and $S$ denotes their strength in such a way, that at the reference point $y_r=6$ Mm magnetic field strength equals $B=2$ Gs. Note that term "magnetic arcade" typically refers to an active region structure; for quiet Sun described magnetic configuration may represent a bipolar region of network magnetic field. The magnitude of the magnetic field, $B$, reaches $100\;$G at the foot points of the magnetic structure and magnetic field lines are essentially vertical there. Higher up, $B$ declines with height. Between the foot points of the magnetic arcade, magnetic field lines become horizontal. Initial magnetic configuration is shown on Fig.~1 (top). The bottom panel of Fig.~1 illustrates the resulting spatial profile of the plasma beta given by 
\begin{equation}
    \beta (x,y) = \frac{p_{\rm i}+p_{\rm n}}{B^2/2\mu_{0}} \, .
\end{equation}

\subsection{Numerical experiments}
We perturb the magnetohydrostatic equilibrium by a very small random signal in the ion and neutral velocities to seed convection below the bottom of the solar photosphere, which is located at $y=0\;$Mm. 
As a result of this initial perturbation, the convection cells start forming to be already discernible after about $10\;$min of the physical time, and fully-developed convection occurs after about $1\;$h. 


To sustain the ongoing convection, the problem of mass and energy losses is fixed in two ways. A plasma inflow at the bottom boundary is implemented in its vertical velocity component equal to $0.3\;$km s$^{-1}$. This allows us to compensate the outflowing plasma mass and its energy losses. 
Additionally, the convection zone ($y<0$ Mm) is heated by implementing the extra source term in the ion energy equation, which overwhelms the radiative cooling term by 10\% there. Such source term mimics the plasma heating which results from the convection zone.
Both these values are calculated by assuming that the incoming flux and heating from below the simulation region are supposed to compensate the mass and energy losses and allow us to sustain an atmospheric quasi-equilibrium \citep{Wojcik2020,Murawski2020}.

Figure~2 illustrates the magnetic solar arcade perturbed by ion MAG and neutral acoustic-gravity waves excited by the ongoing solar convection and granulation, at $t=3800$ s, $t=4000$ s, and $t=4200$ s (from top to bottom). 
The colour-map on the left panels shows the logarithm of the ion temperature, $T_{\rm i}$, expressed in K, 
while the solid lines represent magnetic field lines. 
Note the well-developed granulation pattern which is operating in and below the photosphere, located approximately between $y=0\;$Mm and $y= 0.5\;$Mm. 
Characteristic features of this granulation such as turbulence with strong downdrafts and weak upflows as well as a diversity of waves and flows are well seen. 
From Fig.~2 we infer that dense chromospheric plasma fills the structure of the magnetic solar arcade. 
On both sides of this structure, namely at the foot points of the magnetic arcade, located at $x\cong-5.12$ and $x\cong5.12\;$Mm,  
magnetic field lines form flux tubes. 
Note well-developed magnetic blobs rising from below the photosphere. 
The right panels of Fig.~2 illustrate the corresponding ion-neutral drift, $|V_{\rm i}-V_{\rm n}|$, expressed in km s$^{-1}$ at the same three instants of time. As expected, both fluids are strongly coupled in the photosphere, while they decouple in the chromosphere. This decoupling is especially well seen in the magnetic structure of the arcade, where the ion-neutral drift reaches its maximum value of tens of meters per second.
The magnetic field, mostly horizontal here, does not directly affect the propagating neutral acoustic-gravity waves, while the ion magnetoacoustic-gravity waves propagating upwards do encounter its influence. A part of the energy carried by these waves is thermalized by the process of ion-neutral collisions \cite[e.g.][]{Kuzma2019}. 
Higher up, in the solar corona, the ion-neutral drift loses its importance as plasma is in general fully-ionized there. 

\begin{figure}
\begin{center}
\includegraphics[width=1.0\linewidth]{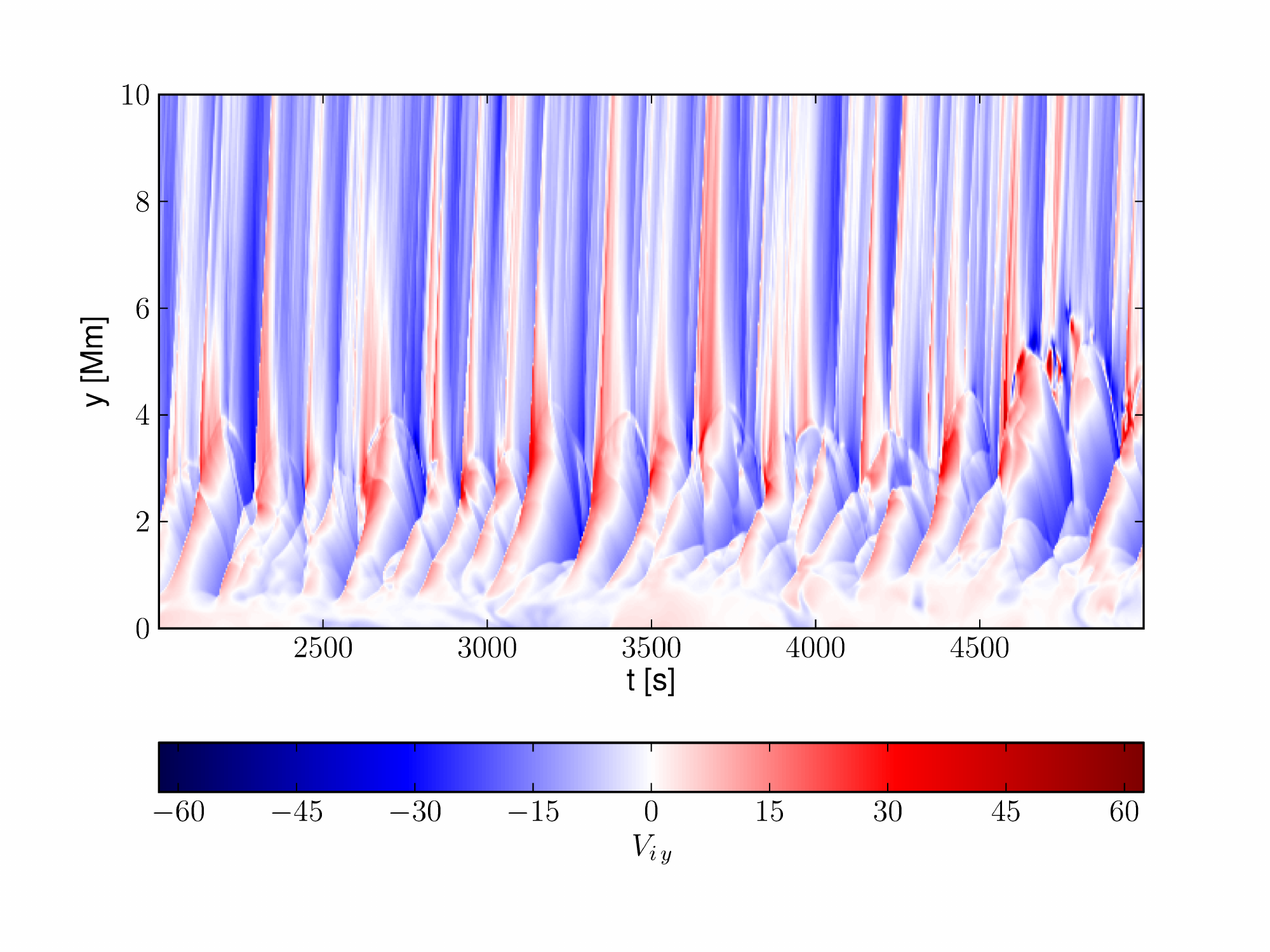}
\end{center}
\vspace{-1cm}
\caption{\small Time-distance plot for $V_{{\rm i}\, y}(x=0,y,t)$, given in units of $1\;$km s$^{-1}$. 
}
\label{fig:timedistance}
\end{figure}
\begin{figure}
\begin{center}
\includegraphics[width=1.0\linewidth]{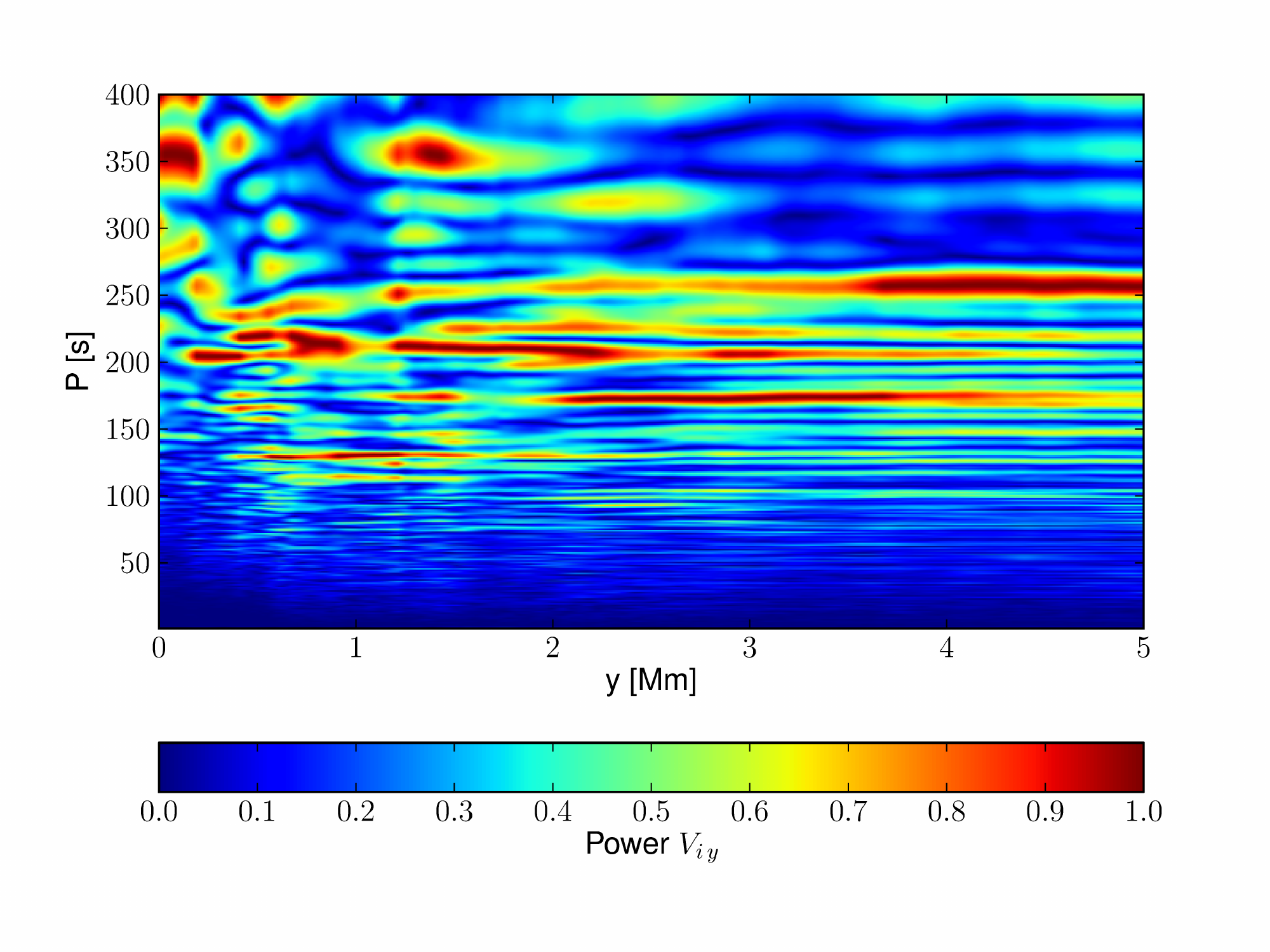}
\end{center}
\vspace{-1cm}
\caption{\small Fourier power spectrum (in arbitrary units) of wave-period $P$ associated with the $V_{i\,y}(x=0,y,t)$ vs height $y$ in the solar atmosphere. 
}
\label{fig:Pvsy}
\end{figure}

Figure~\ref{fig:timedistance} displays time-distance plot for vertical component of ion velocity collected at $x= 0\;$Mm along $y-$direction.
As at this location magnetic field lines are essentially horizontal, $V_{i\,y}(x=0,y,t)$ corresponds to mainly MAWs which experience cut-off period $P_k=230\;$s, studied analytically, by \cite{Stark_1993} and numerically by \cite{Kraskiewicz_2019}. 
Within the MHD model \cite{Kraskiewicz_2019} showed that a monochromatic driver with its period $300\;$s excites fast MAWs which experience their wave-period falling off to about $200\;$s at $y\approx1\;$Mm. 
The oscillations with $P=200\;$s propagate essentially freely through the chromosphere reaching the corona with the same period; see Fig.~5 in \cite{Kraskiewicz_2019}. 
The essentially horizontal magnetic field above the magnetic arcade, we observe plasma outflows with vertical plasma velocity up to $60\;$km s$^{-1}$. 
These outflows are significantly slower than fast plasma outflows in vertical magnetic field case \citep[]{Wojcik2019}. 

Figure~\ref{fig:Pvsy} illustrates the Fourier power spectrum of wave-period $P$ associated with ion vertical velocity, collected in time along the vertical line, given by $x=0\;$Mm. 
Note that for $x=0\;$Mm magnetic field lines are essentially horizontal, and $P$ is close to $227\;$s for all heights within the range of $1\,{\rm Mm}<y<2\,{\rm Mm}$. 
Higher up, for $y>2.1\;$Mm, the signal associated with $P\approx 227\;$s decays and the dominant wave-period is associated with $P\approx 167\;$s. 

\begin{figure}
\begin{center}
\includegraphics[width=0.99\linewidth]{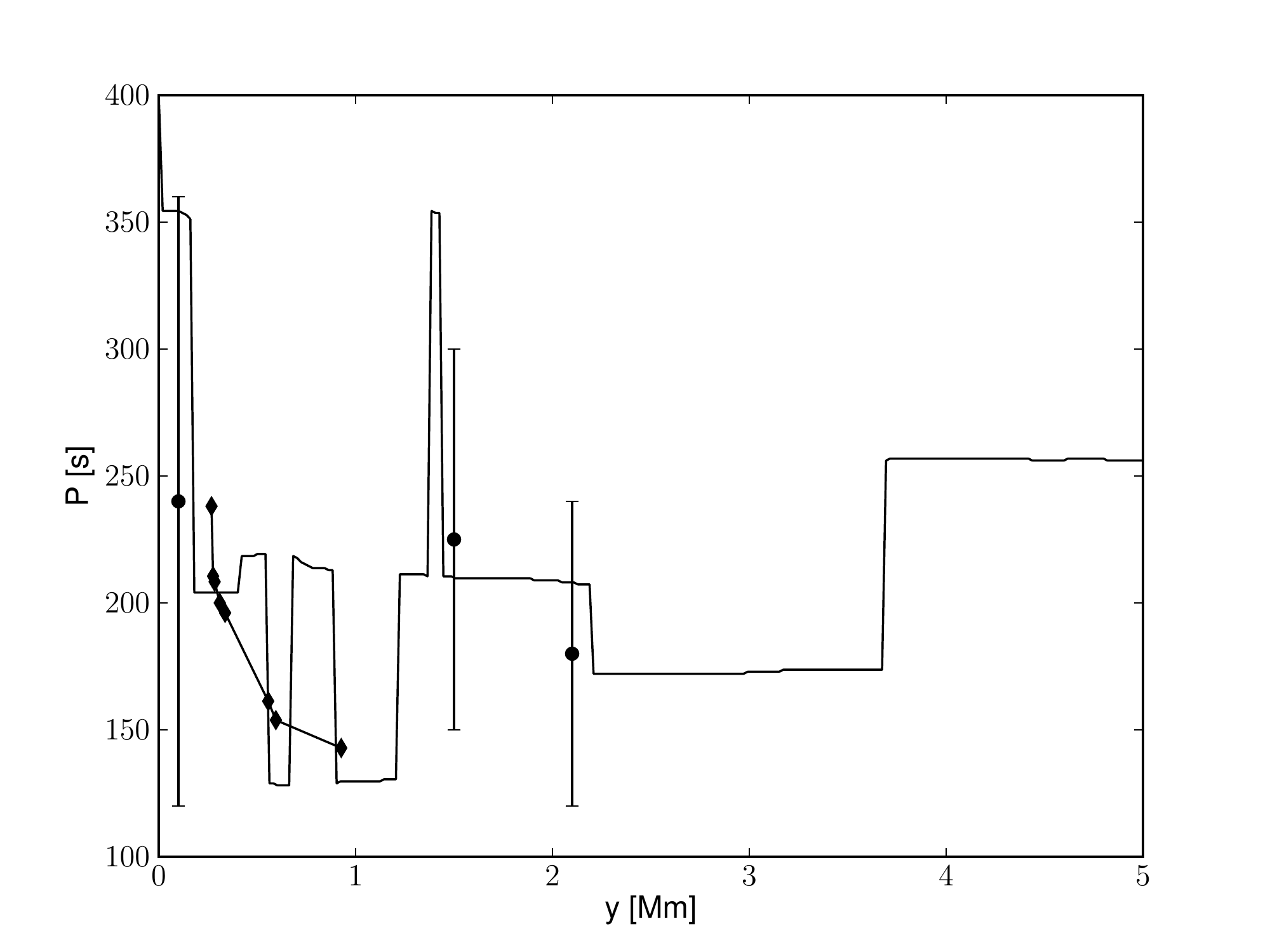}
\end{center}
\vspace{-0.5cm}
\caption{\small Main wave-periods evaluated from Fourier power spectrum for the numerically obtained $V_{i\,y}(x=0,y,t)$ (black solid line) vs height. 
The diamonds and dots show the observational data by \cite{2016ApJ...819L..23W} and \cite{Kayshap2018}, respectively. 
On the latter, error bars represent the range of periods detected by these observations at three given heights.
}
\label{fig:Pvsy-obser}
\end{figure}
Figure~\ref{fig:Pvsy-obser} shows the dominant wave-periods obtained from the Fourier power spectrum for the vertical component of the velocity of the ions and the observational data collected by \cite{2016ApJ...819L..23W} and \cite{Kayshap2018}. 
We infer that the wave-period $P$ follows the observational data, and conclude that the description of the solar atmospheric plasma in the framework of two-fluid equations for ions and neutrals demonstrates a good agreement between numerically detected periods of the waves, which are generated by the solar granulation and the observational data. 

\begin{figure}
\begin{center}
\includegraphics[width=0.99\linewidth]{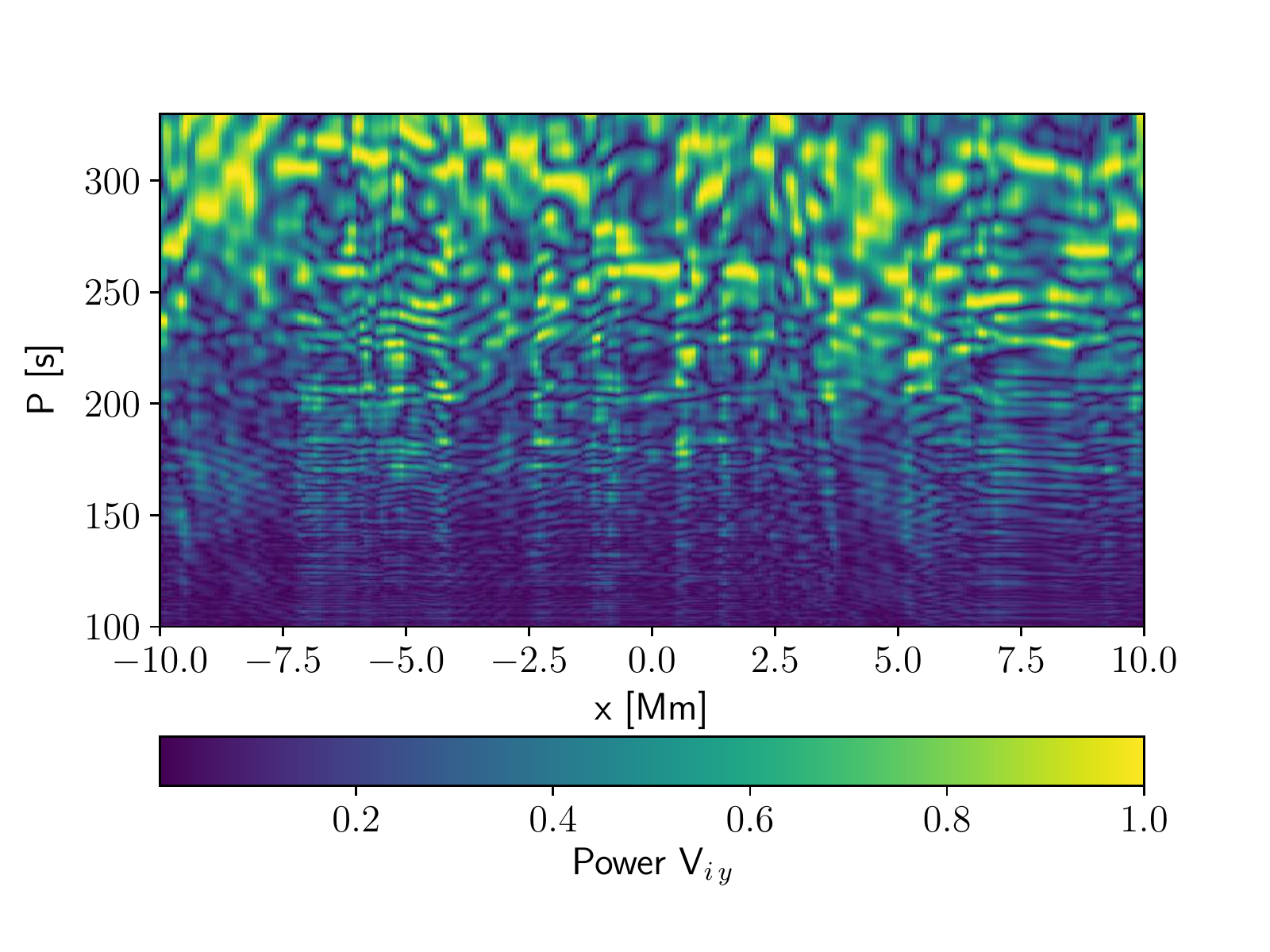}\vspace{-1.6cm}
\includegraphics[width=0.99\linewidth]{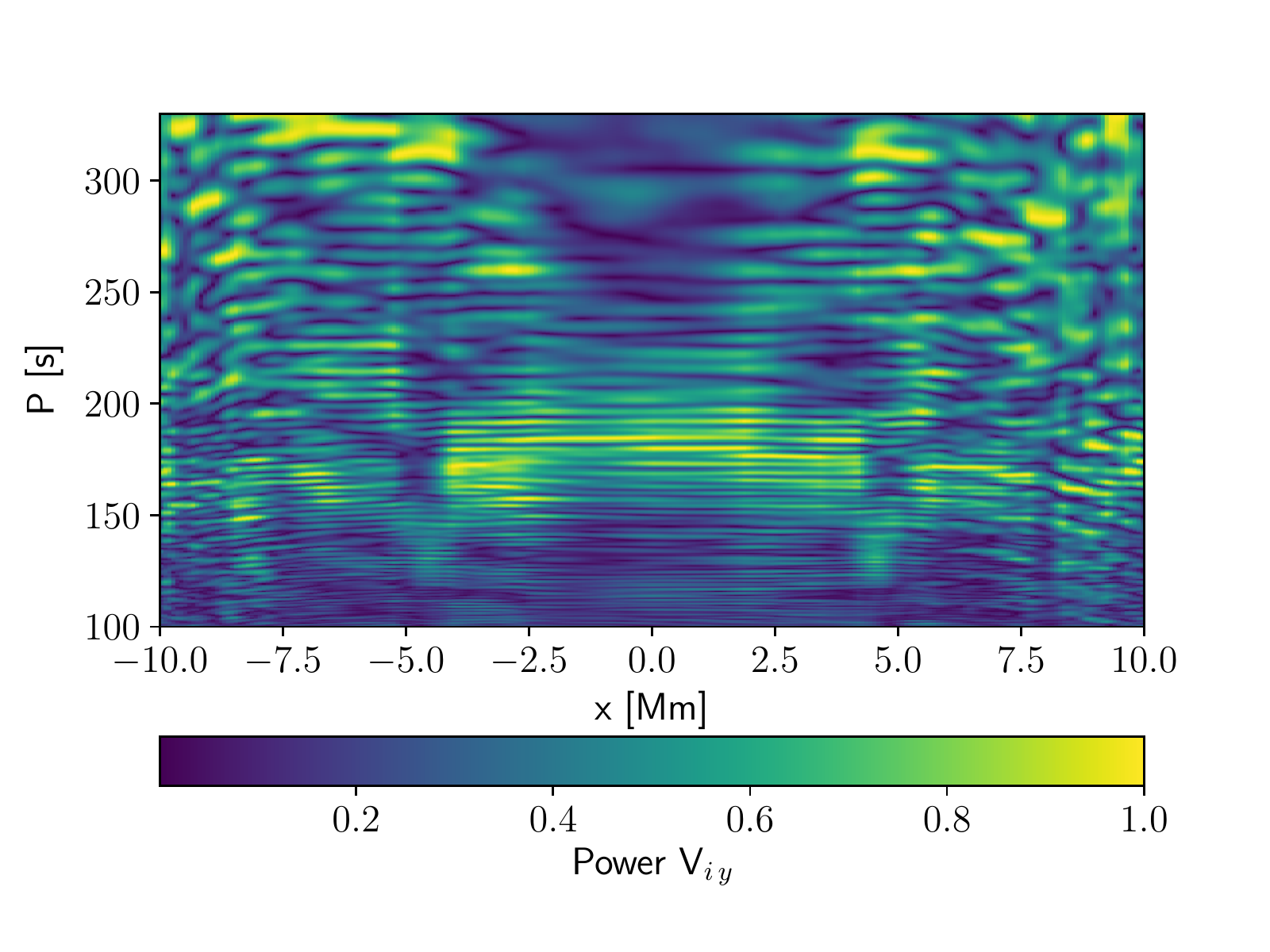}
\end{center}
\vspace{-1cm}
\caption{\small Fourier power spectrum of $P$ (in arbitrary units) associated with ion vertical component of velocity, evaluated at $y=0.25\;$Mm (top) and $y=3.5\;$Mm (bottom), vs horizontal distance $x$ in the solar atmosphere.
}
\label{fig:Pvsx}
\end{figure}

Figure~\ref{fig:Pvsx} displays Fourier power spectrum of the ion velocity, collected along the horizontal lines, given by $y=0.25\;$Mm (top) and $y=3.5\;$Mm (bottom). We infer that the ongoing convective motions below the photosphere generate a broad range of wave-periods that can be observed at the top of the photosphere; the obtained Fourier power shows its maximum between the wave-periods $250\;$s and $350\;$s. While propagating upwards, these waves are subject to complex interactions: mode-coupling, refractions through the inhomogeneous atmosphere, real physical absorption and conversion of wave power (e.g.\ \citealp{Khomenko2018}). Above the foot-points of the arcade, additional shorter wave-periods are generated 
of about $180\;$s and $220\;$s. Above the main loop of the solar arcade, which is for $-5.12\;$Mm$<x<5.12\;$Mm, where mostly horizontal magnetic field lines significantly interact with ion waves, the main spectral power drops to the period of about $180\;$s and longer wave-periods are essentially non-existent.


%
\section{Summary}
We performed 2D numerical simulations of two-fluid ion magnetoacoustic-gravity and neutral acoustic-gravity waves, which are generated by the solar granulation present in low regions of the solar atmosphere. 
This atmosphere is partially-ionized with dynamic ions+electrons and neutrals treated as separate fluids (e.g.\ \citealp{Oliver2016,Ballester2018}), and it is assumed to be initially permeated by a magnetic arcade. The fluids are described by magnetohydrodynamic (for ions+electrons) and hydrodynamic (for neutrals) equations, which are coupled by collision terms, and contain radiation terms corresponding to optically thick and thin cooling \citep{Abbett_2011}. 
The solar granulation, which is generated in the system as a result of convective instability, excites various waves \citep[e.g.][]{Vigeesh2017} 
that are subject of the interaction with magnetic field embedded in the stratified solar atmosphere, cut-off frequencies and an enhancement in high-frequency acoustic power in the solar photosphere and the chromosphere surrounding magnetic active regions.  Moreover\ \cite{Rijs2015,Rijs2016} reported the formation of an acoustic halo caused by MHD mode conversion through regions of moderate and inclined magnetic fields. This conversion type is most efficient when high frequency waves from below intersect magnetic field lines at a large angle.

The main finding of this paper is that in the photospheric regions of highly turbulent and strongly coupled plasma 
the main period of excited waves is close to $300\;$s, 
while in the chromosphere,  where both fluids decouple, situation differs between regions outside and inside of the solar magnetic arcade. It is shown, that in such magnetic structures a strong ion-neutral drift is present, while outside of these structures, ions and neutrals remain essentially coupled.  In addition, refraction and subsequent reflection of magneto-acoustic waves around the regions of $\beta=1$ \citep{Rajaguru2013} significantly alters the wave propagation in this model. 
The waves become evanescent as a result of strong reflection in inhomogeneous solar atmosphere, they decay with height and may transfer their energy into the waves whose wave-periods are close to $3\;$min, which is seen above the solar arcade, while longer wave periods are propagating outside of the structure of the arcade. This is in agreement with findings of \cite{Rajaguru2019}, who showed that less inclined magnetic field elements in the quiet-Sun channel a significant amount of waves of frequency lower than the theoretical minimum acoustic cutoff frequency to the upper atmospheric layers. This is also in a good agreement with the observations reported by \cite{2016ApJ...819L..23W} and \cite{Kayshap2018}. 

The presented results have important implications to the wave energy transfer in the solar atmosphere and its local wave heating, which still remains an unsolved problem of solar physics. In recent work \citet{Fleck2021} compares 10 different 3D magnetoconvection simulations from 4 different codes, namely Bifrost, CO5BOLD, MANCHA3D, and MURaM, under a variety of approximations, and describes waves that are generated within them, and how they vary with height. They found considerable differences between the various models. The height dependence of wave power, in particular of high-frequency waves, varies by up to two orders of magnitude, and the phase difference spectra of several models show unexpected features, including
$ \mp180^{\circ}$ phase jumps.  
Overall several two-fluid effects included in our model will need separate dedicated studies using less complicated and idealized modelling techniques that will allow further comparison and verification. 

\subsubsection*{Acknowledgements}
This work was supported through the projects of National Science  Centre (NCN), Poland, grant nos. 2017/25/B/ST9/00506, 2017/27/N/ST9/01798, 
2020/37/B/ST9/00184, and C14/19/089  (C1 project Internal Funds KU Leuven), G.0D07.19N  (FWO-Vlaanderen), SIDC Data Exploitation (ESA Prodex-12). This project (EUHFORIA 2.0) has received funding from the European Union’s Horizon 2020 research and innovation programme under grant agreement No 870405.

\bibliographystyle{aa} 
\bibliography{draft.bib}
\end{document}